\documentclass[11pt,a4paper]{article}
\usepackage{jcappub}

\newcommand\nice[1]{#1}    \newcommand\subm[1]{}   % format ``pretty print''
\nice{\newcommand\dosingle[1]{#1}  \newcommand\dodouble[1]{ }} % for single spacing
\subm{\newcommand\dosingle[1]{ }   \newcommand\dodouble[1]{#1}} % for double spacing

\usepackage{graphicx}  % graphicx rather than epsf 
\usepackage{dcolumn}% Align table columns on decimal point
\usepackage{bm}% bold math

\newcommand\mystampdothestamp[1]{}  %%private:mystamp-preamble

\usepackage{amsfonts}
 \usepackage{hyperref}
\nice{ 

\providecommand{\url}[1]{\href{#1}{#1}}
\providecommand{\newblock}{}  % maybe not needed?
}

\providecommand{\adsurl}[1]{}

\newcommand\SSS{Sect.~}
\providecommand\apj{ApJ}                 % {Ap. J.}
\providecommand\aj{AJ}                 % {Ap. J.}
\providecommand\apjl{ApJL}                 % {Ap. J.}
\providecommand\apjs{ApJSupp}                 % {Ap. J.}
\providecommand\aap{A\&A}            % {A. \& A.} 
\providecommand\mnras{MNRAS}
\providecommand\cqg{CQG}
\providecommand\grg{Gen. Rev. Grav.} %% General Relativity and Gravitation

\providecommand\prd{Phys.~Rev.~D}
\providecommand\physrep{Phys. Rep.}
\providecommand\jcap{JCAP}
\providecommand\advspaceres{Adv. Space Res.}

\newcommand\mycaptionfont{}

\newcommand\fvir{f_{\mathrm{vir}}}
\newcommand\deltavir{\delta_{\mathrm{vir}}}
\newcommand\OmQD{\Omega_{\cal Q}^{\cal D}}
\newcommand\OmWD{\Omega_{\cal W}^{\cal D}}
\newcommand\OmRD{\Omega_{\cal R}^{\cal D}}

\providecommand\agt{\,\lower.6ex\hbox{$\buildrel >\over \sim$} \, }
\providecommand\alt{\,\lower.6ex\hbox{$\buildrel <\over \sim$} \, }
\newcommand\hMpc{{$h^{-1}$~Mpc}}
\newcommand\hGpc{{$h^{-1}$~Gpc}}

\newcommand\Omm{\Omega_{\mathrm{m}}}%%% EDITOR modify as desired 
\newcommand\Ommzero{\Omega_{\mathrm{m}0}}%%% EDITOR modify as desired 

\newcommand\Ommeff{\Omega_{\mathrm{m}}^{\mathrm{eff}}}

\newcommand\OmReff{\Omega_{\cal{R}}^{\mathrm{eff}}}
\newcommand\chieff{\chi}
\newcommand\aeff{a_{\mathrm{eff}}}
\newcommand\dotaeff{\dot{a}_{\mathrm{eff}}}
\newcommand\ddotaeff{\ddot{a}_{\mathrm{eff}}}

\newcommand\Heff{H^{\mathrm{eff}}}

\newcommand\RCeff{R_{\mathrm{C}}^{\mathrm{eff}}}
\newcommand\dLeff{d_{L}^{\mathrm{eff}}}
\newcommand\dVdzeff{\diffd {V}^{\mathrm{eff}}/\diffd z}
\newcommand\dVdzeffmath{\frac{\diffd {V}^{\mathrm{eff}}}{\diffd z}}
\newcommand\qdeceleff{q^{\mathrm{eff}}}
\newcommand\diffd{\mathrm{d}}
\newcommand\vinfall{v_{\mathrm{infall}}}
\newcommand\Dvoid{D_{\mathrm{void}}}

\newcommand\Hbg{H_0^{\mathrm{bg}}}
\newcommand\supbg{^{\mathrm{bg}}}
\newcommand\Hpeculiarcomov{H_{\mathrm{pec}}^{\mathrm{com}}}

\newcommand\fracdistmod{f_{m-M}}

\newcommand{\CD}{{\cal D}}
\newcommand{\CE}{{\cal E}}
\newcommand{\CF}{{\cal F}}
\newcommand{\CQ}{{\cal Q}}
\newcommand{\CR}{{\cal R}}
\newcommand{\CM}{{\cal M}}
\newcommand{\average}[1]{\left\langle #1 \right\rangle_\CD}
\newcommand{\averageM}[1]{\left\langle #1 \right\rangle _{{\cal M}}}
\newcommand{\averageE}[1]{\left\langle #1 \right\rangle _{{\cal E}}}

\newcommand\prerefereechanges[1]{#1}    

\newcommand\postrefereechangesI[1]{#1}

\newcommand\postrefereechangesII[1]{#1}

\title{Virialisation-induced curvature as a physical explanation for dark energy}

\newcommand\TCfAaddress{Toru\'n Centre for Astronomy, 
  Faculty of Physics, Astronomy and Informatics,
  Nicolaus
  Copernicus University, ul. Gagarina 11, 87-100 Toru\'n, Poland}
\newcommand\CRALaddress{Universit\'e de Lyon, Observatoire de Lyon,
Centre de Recherche Astrophysique de Lyon, CNRS UMR 5574: Universit\'e Lyon~1 and \'Ecole Normale Sup\'erieure de Lyon, 
9 avenue Charles Andr\'e, F--69230 Saint-Genis-Laval, France}

\author[a,b,1]{Boudewijn F. Roukema \note{Affiliation b: during visiting lectureship.}}
\author[a,b,2]{Jan J. Ostrowski \note{Affiliation b: during long-term visit.}}
\author[b]{Thomas Buchert}
\affiliation[a]{\TCfAaddress}
\affiliation[b]{\CRALaddress}
\date{\today}

\abstract{
  {The geometry of the dark energy and 
  cold dark matter dominated
     cosmological model ($\Lambda$CDM) is commonly assumed to be given by a
    Friedmann--Lema\^{\i}tre--Robertson--Walker (FLRW) metric, i.e. it
    assumes homogeneity in the comoving spatial section. 
    {\protect\postrefereechangesI{The homogeneity assumption fails
      most strongly at (i) small distance scales and
      (ii) recent epochs, implying that the FLRW approximation
      is most likely to fail at these scales.}}}
  {We use the virialisation fraction to quantify (i) and (ii), which
    approximately coincide with each other 
    on the observational past
    light cone. For increasing time, the virialisation fraction
    increases above 10\% at about the same redshift ($\sim 1$--$3$)
    at which $\Omega_\Lambda$ grows above 10\% ($\approx 1.8$).}
  {Thus, instead of non-zero $\Omega_\Lambda$, we propose an
    approximate, general-relativistic correction to the
    matter-dominated ($\Omm =1, \Omega_\Lambda=0$), homogeneous
    metric (Einstein--de~Sitter, EdS).}
  {{A low-redshift effective matter-density parameter 
      of $\Ommeff(0) = 0.26 \pm 0.05$ is inferred.
      Over redshifts $0 < z < 3$, 
      the distance modulus of the virialisation-corrected
      EdS model approximately matches the 
      $\Lambda$CDM distance modulus.}}
  {This rough approximation assumes ``old physics'' (general
    relativity), not ``new physics''. 
    {Thus, pending more {detailed}
      calculations, we strengthen the claim that ``dark
    energy'' should be considered as an artefact of emerging average
    curvature in the void-dominated Universe, via a novel approach
    that quantifies the relation between virialisation and average
    curvature evolution.}}
}
\begin{document}

%\maketitle

\mystampdothestamp{}

\maketitle % IOP positioning of abstract

%\dodouble{\clearpage} %% otherwise first text section is funny

%%%%%%%%%%%%%%%%%%%%%%%%%%%%%%%%%%%%%%%%%%%%%%%%%%%%%%%%%%%%%%%%%%%
%% Figure section. Defined early so that position in output can
%% be easily moved towards earlier pages if LaTeX wants to put
%% them all at the end...
%%%%%%%%%%%%%%%%%%%%%%%%%%%%%%%%%%%%%%%%%%%%%%%%%%%%%%%%%%%%%%%%%%%

\newcommand\fDEvsvirial{
  \begin{figure}  % [ht]
    \centering
    \includegraphics[width=9cm]{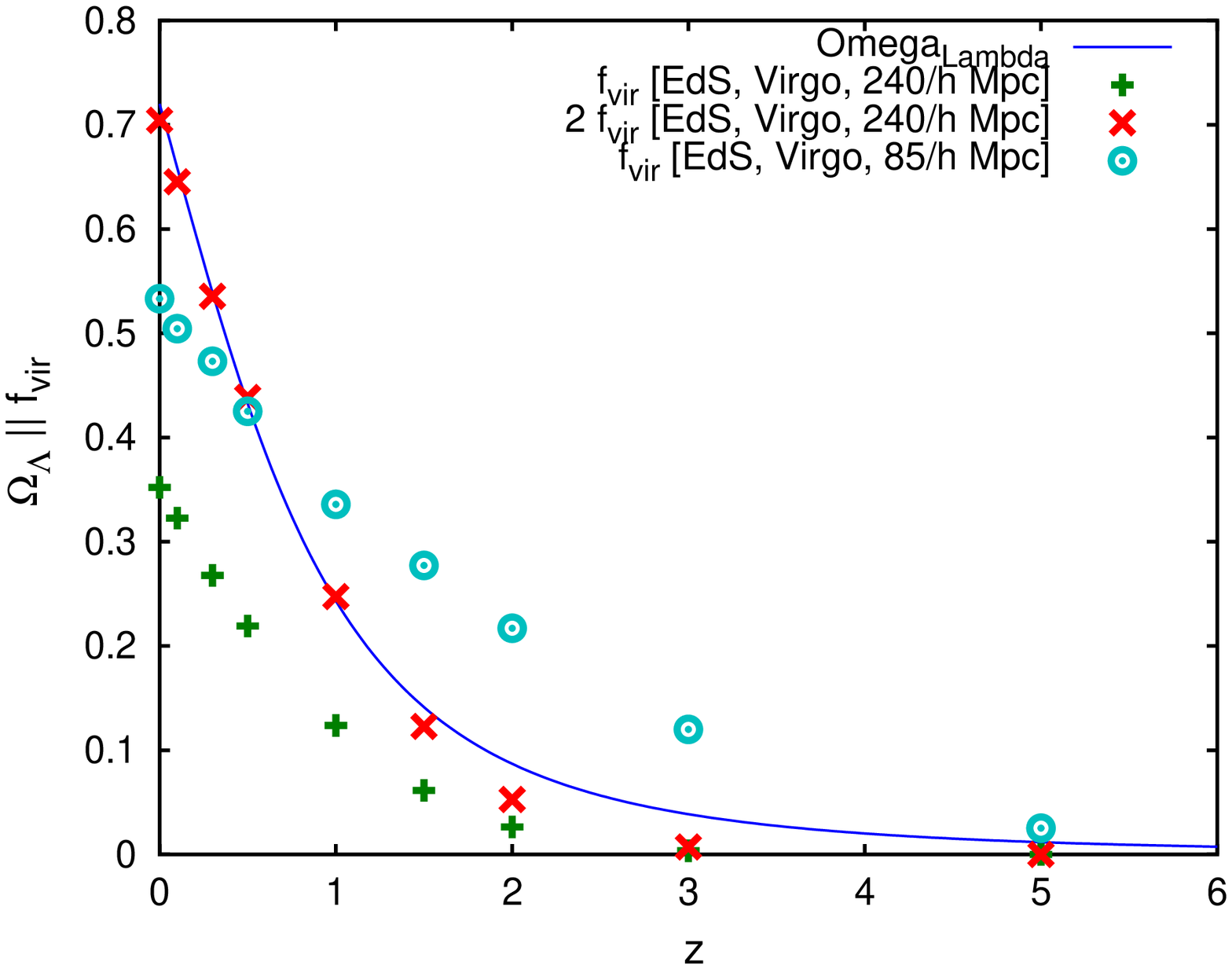}
    \caption[]{ 
      \mycaptionfont (colour online) 
      Redshift evolution of $\Omega_\Lambda$ 
      [Eq.~(\ref{e-lambda-z})] in the $\Lambda$CDM model compared to 
      that of the fraction of virialised non-relativistic 
      (non-baryonic plus baryonic) matter, $\fvir(z)$, 
      in EdS $256^3$-particle
      Virgo Consortium  \protect\citep{Jenkins98Virgo,Thomas98Virgo} 
      $N$-body simulations of 
      box sizes as labelled
      (see Sect.~\protect\ref{s-N-body}).
      For comparison with $\Omega_\Lambda$,
      $2\fvir$ is also shown for the 240~{\hMpc} simulation.
      \label{f-DEvsvirial}
    }
  \end{figure}
} % of \def\fDEvsvirial

\newcommand\tNbody{
  \begin{table*}
    \caption{\mycaptionfont 
      {$N$-body simulation estimates of 
      $\fvir(z)$ (\protect\ref{e-defn-fvir})}
      \label{t-N-body}}
    $$\begin{array}{c rrrrr rrrrr} \hline
      z & 10.0  & 5.0  & 3.0  & 2.0  & 1.5  & 1.0  & 0.5  & 0.3  & 0.1  & 0.0  \\
      \fvir & 0  & 1.4 \times 10^{-5} & 0.0036  & 0.026  & 0.061  & 0.12  & 0.22  & 0.27  & 0.32  & 0.35  \\
      \hline
    \end{array}$$ \\
  \end{table*}
}  %\newcommand\tNbody

\newcommand\fommeff{
  \begin{figure}  % [ht]
    \centering
    \includegraphics[width=7cm]{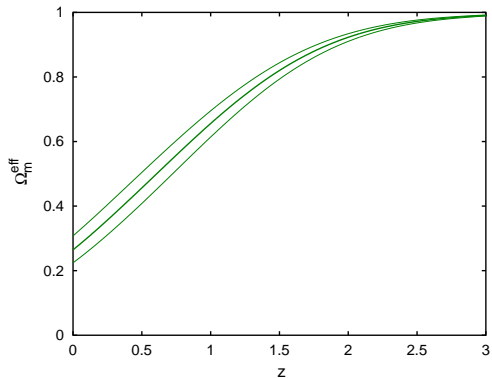}
    \caption[]{ \mycaptionfont %% (colour online) % no point for just one curve
      Effective zero redshift matter density 
      parameter {$\Ommeff$} (\protect\ref{e-ommeff-approx})
      in the virialisation approximation.
      {The upper, central, and lower curves correspond
        to $\Hpeculiarcomov(0) = 33, 36,$ and $39$~km/s/Mpc, respectively
        [see (\protect\ref{e-obs-Hpeccomov})].}
      \label{f-ommeff}
    }
  \end{figure}
} % of \def\fommeff

\newcommand\fHeff{
  \begin{figure}  % [ht]
    \centering
    \includegraphics[width=7cm]{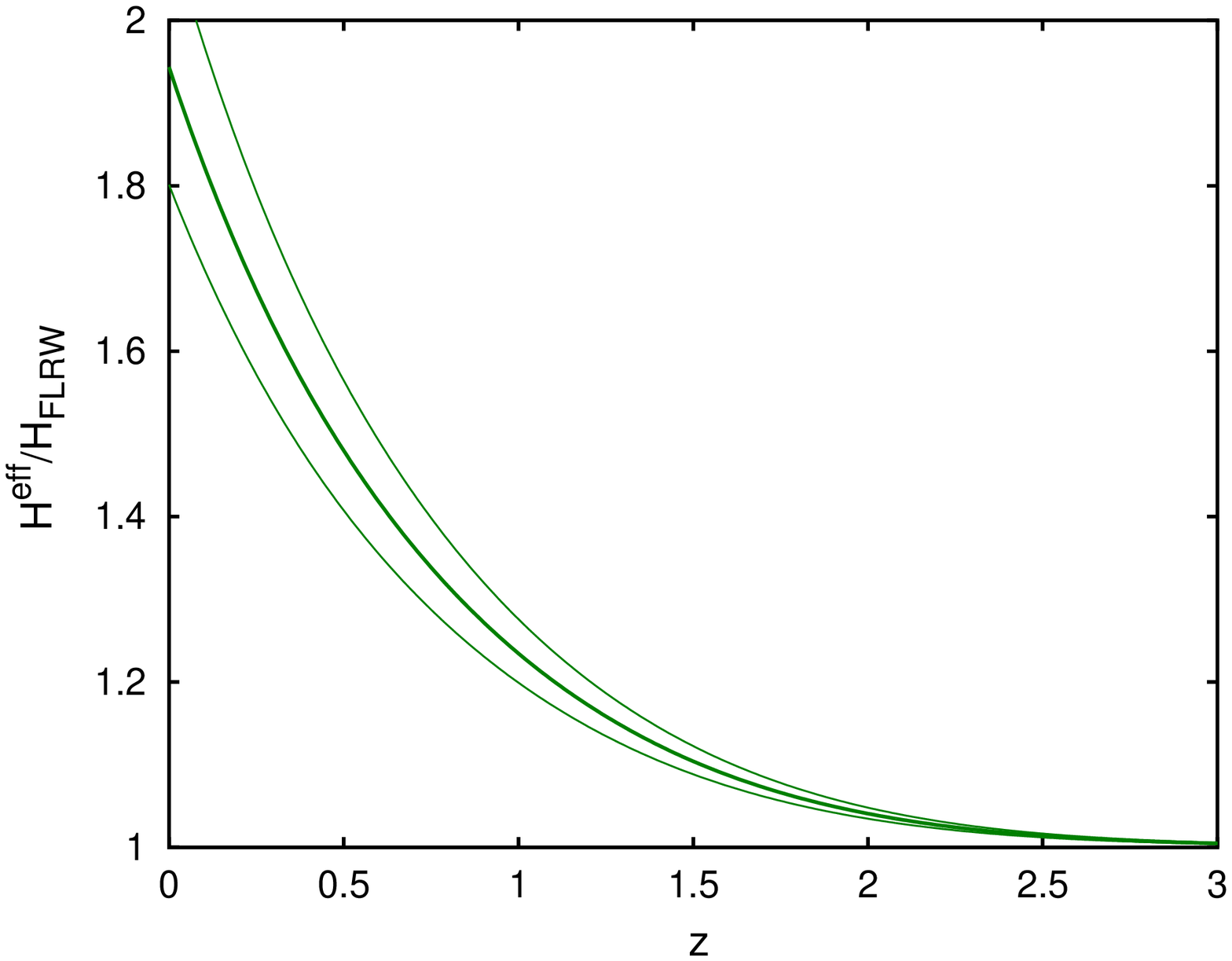}
    \caption[]{ \mycaptionfont %% (colour online)  % no point for just one curve
      {Ratio of effective to background
        expansion rates $\Heff(z)/H(z)$, using
        (\protect\ref{e-defn-Heff}),
        (\protect\ref{e-H-EdS}), and
        (\protect\ref{e-Hbg-assumption}).}
      {The upper, central, and lower curves correspond
        to $\Hpeculiarcomov(0) = 39, 36,$ and $33$~km/s/Mpc, respectively.}
      \label{f-Heff}
    }
  \end{figure}
} % of \def\fHeff

\newcommand\fRCeff{
  \begin{figure}  % [ht]
    \centering
    \includegraphics[width=7cm]{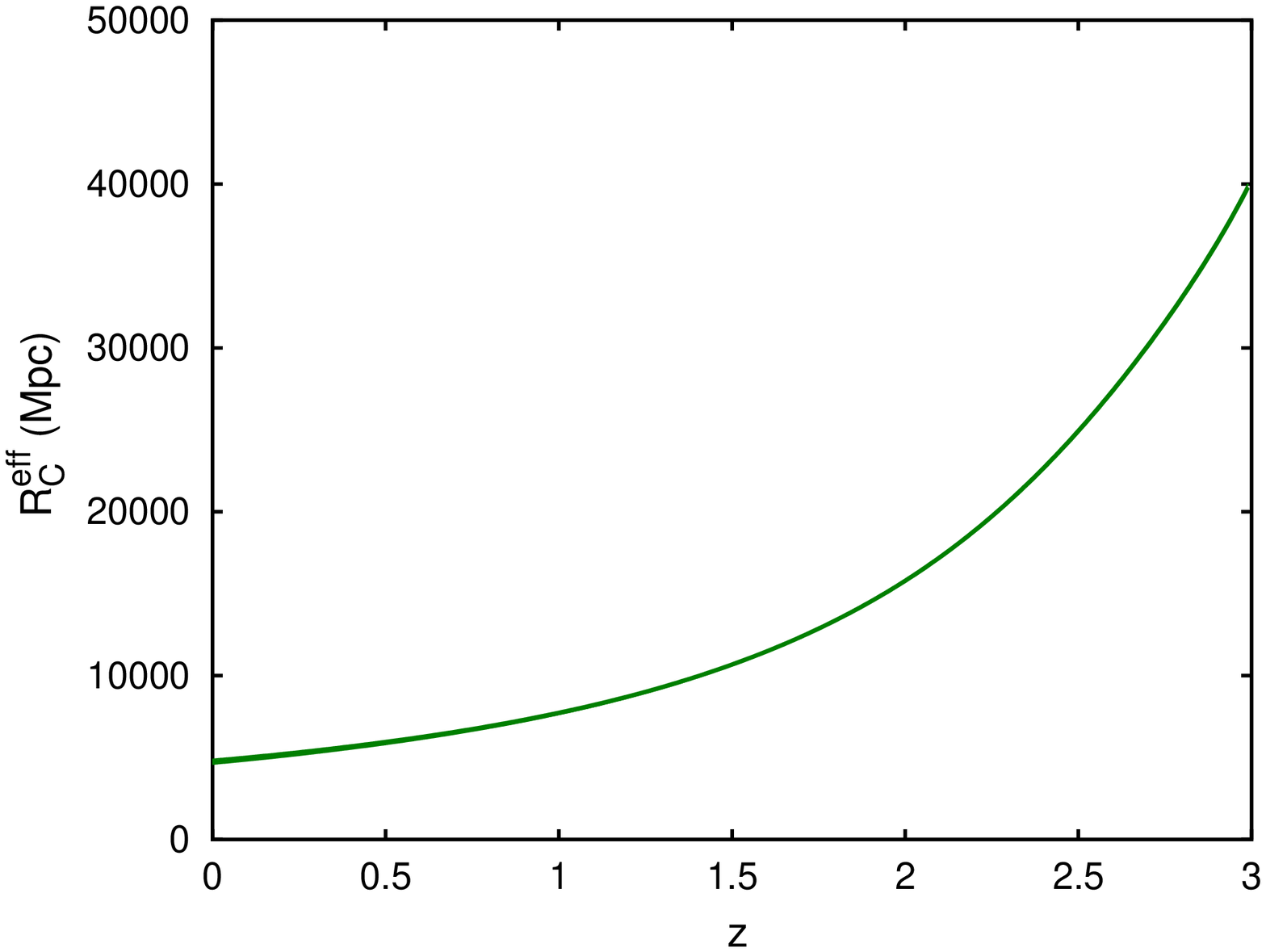}
    \caption[]{ \mycaptionfont %% (colour online)  % no point for just one curve
      Effective {comoving} curvature radius 
      $\RCeff$ (\protect\ref{e-defn-RCeff})
      in the virialisation {approximation.}
      {The uncertainty in $\Hpeculiarcomov(0)$ is
        not visible in this plot.}
      \label{f-RCeff}
   }
  \end{figure}
} % of \def\fRCeff

\newcommand\fmetricVF{
  \begin{figure}  % [ht]
    \centering
    \includegraphics[width=9cm]{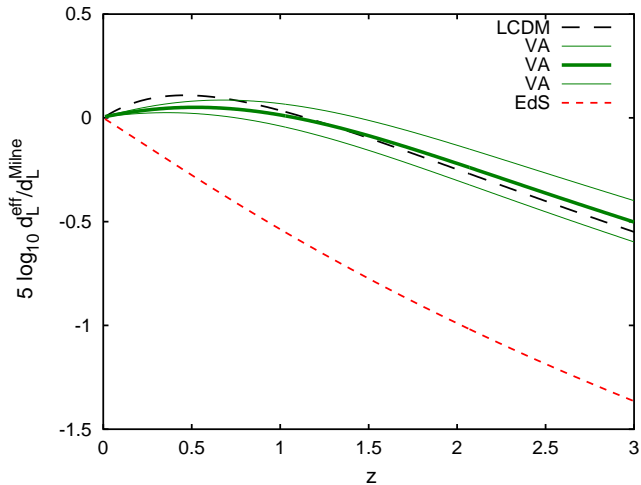}
    \caption[]{ \mycaptionfont (colour online) Distance moduli for the EdS
      (bottom, \protect\postrefereechangesII{short-dashed} curve) and $\Lambda$CDM 
      (\protect\postrefereechangesII{long-dashed curve, top for} $z \alt 0.5$, black online)
      {homogeneous} models, and 
      for the {EdS} 
      virialisation approximation (\SSS\protect\ref{s-method-fvir}; 
      {green online, thick curve 
      for $\Hpeculiarcomov(0)=36$~km/s/Mpc, thin curves for 
      $\Hpeculiarcomov(0)=33$ and $39$~km/s/Mpc}),
      {all} 
      normalised to the Milne model
      ($\Ommzero= 0 =\Omega_{\Lambda0}$). 
      \label{f-metricVF}
    }
  \end{figure}
} % of \def\fmetricVF

\newcommand\fmetricVFfraction{
  \begin{figure}  % [ht]
    \centering
    \includegraphics[width=7cm]{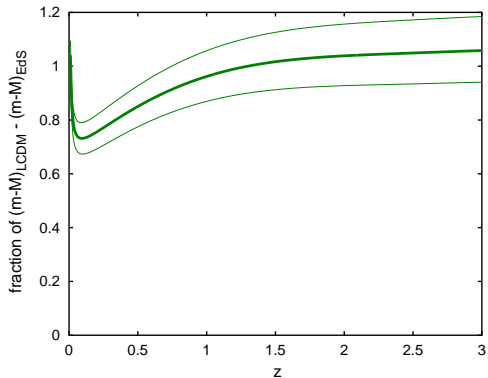}
    \caption[]{ \mycaptionfont %% (colour online)
      Fraction of EdS-to-$\Lambda$CDM
      distance modulus $\protect\fracdistmod$
      (\ref{e-defn-fracdistmod})
      provided by the virialisation 
      {approximation}.
      \label{f-metricVFfraction}
    }
  \end{figure}
} % of \def\fmetricVFfraction

\newcommand\fmetricVFdVdz{
  \begin{figure}  % [ht]
    \centering
    \includegraphics[width=7cm]{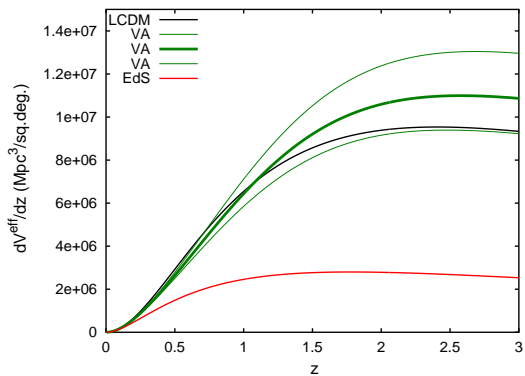}
    \caption[]{ \mycaptionfont (colour online)
      Differential volume element $\dVdzeff$
      (\protect\ref{e-dVdz-eff}) for the EdS (bottom curve) and
      $\Lambda$CDM {(black online) homogeneous models and for the EdS
        virialisation approximation (thick and very thin curves
        for $\Hpeculiarcomov(0) = 39, 36,$ and $33$~km/s/Mpc, from top to bottom, respectively,
        green online),} in comoving Mpc$^3$/sq.deg.
      \label{f-metricVFdVdz}
    }
  \end{figure}
} % of \def\fmetricVFdVdz

\newcommand\fmetricVFqdecel{
  \begin{figure}  % [ht]
    \centering
    \includegraphics[width=7cm]{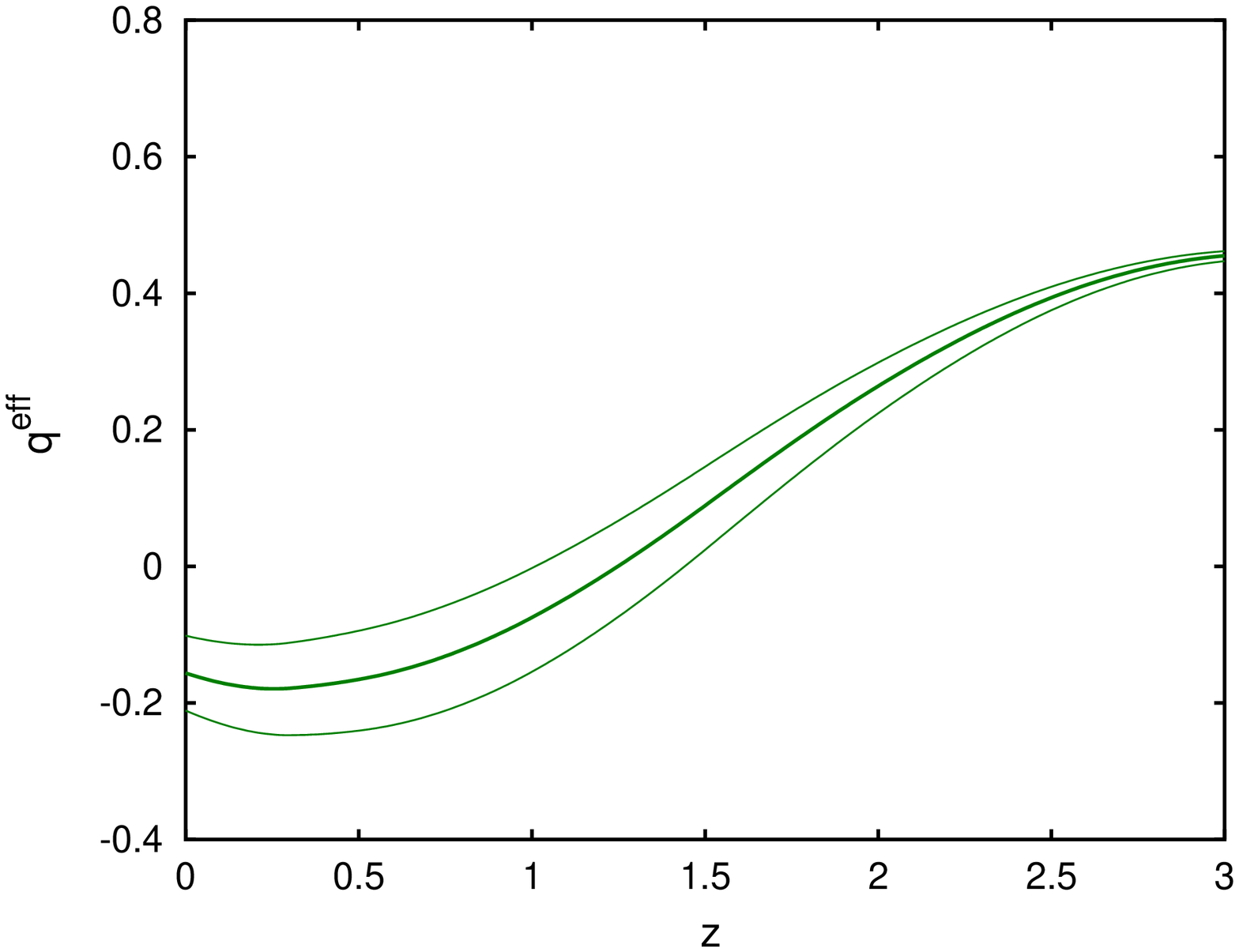}
    \caption[]{ \mycaptionfont %% (colour online)  % no point for just one curve
      Effective deceleration parameter $\qdeceleff$ 
      (\protect\ref{e-defn-q-decel}) for the EdS virialisation approximation
      presented in this {paper}.
      \label{f-metricVFqdecel}
    }
  \end{figure}
} % of \def\fmetricVFqdecel

\section{Introduction}   \label{s-intro}

%context (optional)

It is widely believed that a valid general-relativistic
interpretation of recent {extragalactic observations}, in
particular faint galaxy number counts 
\postrefereechangesI{\citep[e.g.][]{FYTY90,YP95},
gravitational lensing \citep[e.g.][]{ChY97,FortMD97}}, supernovae type Ia
magnitude-redshift relations \citep[e.g.][]{SCP9812,SNeSchmidt98}),
and the Wilkinson Microwave Anisotropy Probe (WMAP) observations of
the cosmic microwave background \citep{WMAPSpergel}, is that the
present-day Universe is dominated by a non-zero dark
energy term, modelled in the simplest case by a cosmological constant with today's 
value 
{$\Omega_{\Lambda0} \approx 0.68$ \citep{PlanckXVIcosmoparam13}}.
This interpretation is a consequence of forcing the 
Friedmann--Lema\^{\i}tre--Robertson--Walker (FLRW) model
\citep{deSitt17,Fried23,Fried24,Lemaitre31ell,Rob35}, an exact,
\postrefereechangesI{locally homogeneous and 
isotropic}
solution of Einstein's equations
onto the 
observational data
\postrefereechangesII{(for local versus global inhomogeneity and/or anisotropy, see \citep[e.g.][and references therein]{Rouk02asr})}.
What are the key assumptions in applying the FLRW
model to the data?

One key assumption is an applied mathematics hypothesis, called the
``Cosmological Principle'' {that we rephrase here in a {weaker form than usual}: 
{a synchronous} space-time foliation of the Universe
should exist according to which each spatial section 
can be approximated by a constant-curvature metric on some assumed large
{\em scale of homogeneity}, and {the} evolution of this metric is given by the homogeneous-isotropic FLRW solution.}   
Since the real Universe is (obviously) inhomogeneous,
a second assumption is required: that 
the cosmic web of filaments, clusters of galaxies and galaxies
themselves induce only small perturbations of the perfectly
homogeneous geometry of the FLRW background.
We may think of the background as a {\em template space-time}, the validity of which should be questioned. 
{(For the question of which background to use and the construction of 
template metrics employed for the interpretation of observations, see 
\cite{Buchert08status,Kolb10cosmbg,Larena09template}.)}

At what times and length scales are these two assumptions most
questionable?
They are least accurate at recent times (redshifts $z \alt 3$)---since galaxies and
large-scale structure have mostly formed recently---and at small ($\ll
c/H_0 = 3$~{\hGpc}) length scales---where structures have had
sufficient time to become non-linear and to be bound as virialised objects. The
assumption that these structures imply small perturbations of the geometry of the background
may be valid for metric perturbations, but it is violated for the derivatives of the
metric, in particular for the intrinsic curvature \cite{BuchEllvElst09}.

On the observational past {light} cone, the recent epoch
and small length scale regimes, at which we {\em expect\/} that the FLRW 
solution may fail, approximately coincide.

In the FLRW model, $\Omega_\Lambda$ varies with time. At what epochs
is $\Omega_\Lambda$ significantly non-zero, assuming that the present-day
value {is
$\Omega_{\Lambda0} = 0.68$?} Dark energy is significantly non-zero
at recent epochs ($z \alt 3$).

Hence, the FLRW model should reasonably be {\em expected\/} to be a
bad approximation on the same scales at which $\Omega_{\Lambda}$
inferred from applying the FLRW model is significantly non-zero. Since
general relativity is well-established empirically
\postrefereechangesI{(especially on stellar system scales
  \citep{Will06LivReview}) and since alternative gravity
  models usually violate the experimentally well-supported strong
  equivalence principle
  \citep[e.g.][]{Hui09fRSEPviol,Deruelle11SEPviol},} the conservative
scientific approach is to assume that non-zero dark energy is an
artefact of trying to apply the FLRW model in domains where it is
expected to fail, unless or until proven otherwise.
%% Concordance Model \citep{CosConcord95} values of the parameters in the
%%metric,

Can the relation between the scales of the expected failure of the
FLRW model and the significantly non-zero values of FLRW-inferred
$\Omega_\Lambda$ be quantified?  An obvious statistic to quantify the
inhomogeneity of the Universe is the fraction of non-relativistic
matter (baryonic and non-baryonic matter together) that is contained
in virialised (gravitationally bound) systems at a given epoch,
$\fvir(z)$. In a flat FLRW model, ignoring radiation density, 
\begin{eqnarray}
  \Omega_\Lambda(z) 
  &=& 1 - \Omm(z) \nonumber \\
  &=& 1 - \frac{\Ommzero H_0^2}{a^3 H(z)^2} 
  = 1 - \frac{\Ommzero}{a^3 \Omega_{\Lambda0} + \Ommzero } \;.
 \label{e-lambda-z}
\end{eqnarray}

%\clearpage %% HACK - ignore
\fDEvsvirial  
% \addtocounter{figure}{3} \fmetricVF  \clearpage %% HACK - ignore
%\clearpage %% HACK - ignore

Figure~\ref{f-DEvsvirial} shows that on a {\em linear\/} redshift scale,
the epoch of virialisation in an 
Einstein-de~Sitter (EdS, $\Ommzero=1, \Omega_{\Lambda0}=0$)
model with a cold dark matter (CDM) initial power spectrum (EdS-CDM)
coincides closely with the epoch at which $\Omega_\Lambda$
becomes non-negligible in the Concordance Model 
(\citep{CosConcord95}, $\Lambda$CDM), with the details depending on the limitations
of the $N$-body simulations (box size, mass resolution, {group-finding} algorithms, etc.).
The EdS-CDM virialisation fraction 
and the FLRW-inferred dark energy $\Omega_\Lambda$ evolve
similarly, to much better than an order of magnitude.
Thus, {\em the FLRW-inferred $\Omega_\Lambda$ dominates the Universe in a way that 
approximately imitates the degree to which the FLRW geometry should fail,}
shown in Fig.~\ref{f-DEvsvirial} for an EdS-CDM model.

How can the virialisation fraction and the FLRW-inferred dark energy
parameter be more precisely related? 
Since virialised matter occupies 
about 1/100-th to 1/200-th of the volume
that it would occupy if distributed according to the FLRW background
density (e.g. App.~A in \citep{LaceyCole93MN}), 
the volume-average of the 3-spatial curvature should become dominated by 
the curvature corresponding to the volume occupied by the remaining
non-virialised matter, 
generating, on average, negative spatial curvature (if the initial
spatial curvature is zero). 

In an EdS model the virialisation overdensity is, {according to a top-hat estimate} 
{using the scalar virial theorem for isolated systems,}
\begin{equation}
  \deltavir = 18 \pi^2 \approx 178\;,
  \label{e-deltavir-EdS}
\end{equation}
and for low density, zero dark energy models it is 
bounded below by $8\pi^2$ \citep{LaceyCole93MN}.
Since the curvature is effectively negative,
the density of the non-virialised matter inferred from assuming zero
curvature is an overestimate. Since matter mostly flows out of voids,
the rate of expansion (Hubble parameter) in the majority of the volume, i.e. in non-virialised
low-density regions, is higher than the background model average expansion
rate. Again, the matter density parameter, inferred from assuming a homogeneous
Hubble parameter, is overestimated in comparison to calculating it
in proportion to the critical density of the non-virialised region itself.
Thus, the degree of negative curvature is underestimated both due to 
curvature itself and due to deviations in the expansion rate, which can be condensed into a single
$\cal X$-matter parameter (see below).

The scalar averaging approach to relativistic, inhomogeneous models of the
Universe 
(e.g. \citep{BKS00,Buchert08status,BuchCarf08,BuchRZA1,BuchRZA2};
\postrefereechangesI{also \citep{Kolb05a,KolbMNR05,Rasanen06superhoriz,Rasanen06GRF,Kolb11FOCUS}})
focuses
on the kinematical backreaction, 
{$\OmQD(z) \equiv -{\cal Q}_{\cal D} / (6H_{\cal D}^2)$}, and the average
curvature parameter 
{$\OmRD(z) \equiv - \langle {\cal R} \rangle_{\cal D} / (6 H_{\cal D}^2)$,} 
where $H_{\cal D}$ is the volume-averaged Hubble rate 
within a spatial compact domain $\cal D$.
The former is an averaged expression of
extrinsic curvature invariants, which can be interpreted kinematically and depend (in
general) on the variance in the expansion rate and the averaged rates of shear and vorticity,
while the latter represents the average
scalar curvature (3-Ricci curvature) over the domain $\cal D$.
\postrefereechangesII{The domain $\CD$ may be any spatial domain, but in the main
  body of this work, it is especially used to refer to a large-scale
  domain that includes both virialised and non-virialised regions, i.e.
  on what we assume is a scale of homogeneity.}
The overall backreaction effect can be condensed into an $\cal X$-matter parameter,
defined $\Omega^{\cal D}_{\cal X} (z): = \OmQD(z)+\OmRD(z)$
{and} obeying the Hamilton constraint 
on the averaged density parameters
\begin{equation}
  \Omm^{\cal D} (z) + \Omega^{\cal D}_{\cal X} (z) + \Omega^{\cal D}_\Lambda 
  {(z)} \;=\;1\;,
  \label{e-hamiltonconstraint}
\end{equation}
where {$\Omega^{\cal D}_\Lambda := \Lambda/(3H_{\cal D}^2)$} is the volume-averaged 
dark energy parameter.
 
Earlier calculations, starting with a homogeneous EdS
background model, suggest that the peculiar curvature parameter, i.e. the deviation of the total averaged $3$-Ricci curvature from a constant-curvature model,
$\OmWD(z): = - {\cal W}_{\cal D} / 6 H_{\cal D}^2$, 
with ${\cal W}_{\cal D}:=\langle {\cal R} \rangle_{\cal D} - 
\postrefereechangesII{6 k_{\CD} / a_{\CD}^2}$, 
has a much stronger 
(\postrefereechangesII{generally at least} 
a factor of about 5) effect than the
kinematical backreaction $\OmQD(z)$ (cf. 
\postrefereechangesII{Figs~3--5 in \citep{BuchRZA2}}).
Thus, the use of virialisation 
to calculate more realistic estimates of the average scalar curvature and its evolution should provide an
approximate, general-relativistic space-time model that is more
accurate than the FLRW model.
This {effective-metric}
 model is not expected to satisfy the 
  Einstein {equation} on any given (recent) time slice
  {\citep{Larena09template}}.
Thus, the effective metric presented below 
(\ref{e-metric-eff}) is
unlikely to be consistent with the Lema\^{\i}tre--Tolman--Bondi (LTB)
model \citep{Lemaitre33,Tolman34,Bondi47}, which has recently been
parametrised against observations 
\citep{CBK10,KolbLamb09}.

\postrefereechangesI{Alternative scalar-averaging,
  inhomogeneous, dark-energy--free approaches to
  ours include the timescape model and the two--FLRW-component
  toy model.
  In the timescape model
  (formerly known as the ``fractal bubble'' model)
  \citep{Wiltshire07clocks,Wiltshire07exact,Wiltshire09timescape},
  late epochs are modelled by a negatively-curved, void-dominated,
  multi-scale, scalar averaging formalism that focuses particularly on
  time calibration.
  The model compares well to the $\Lambda$CDM model
  in fitting supernovae type Ia data \citep{SmaleWilt11SNe},
  has been used to interpret bulk velocity flows on scales up 
  to 65~{\hMpc} from the Local Group \citep{Wiltshire12Hflow},
  \postrefereechangesII{and has been extended to include
    radiation \cite{DuleyWilt13}}.
  The two--FLRW-component toy model assumes that voids and walls 
  can be separately represented by FLRW dust models \citep{BoehmRasan13}.
  Our approach differs from both in that it directly 
  relates two-component scalar averaging \citep{WiegBuch10} to
  the evolution of the virialisation fraction (\SSS\ref{s-formulae-general}),
  and does not assume that either component evolves separately as an
  FLRW model.}

The approximation method proposed here, for an EdS background model, 
is described 
in \SSS\ref{s-method-fvir}.
\postrefereechangesII{The term {\em background model}, in this
  work denoting a high-redshift FLRW model
  extrapolated to low redshift,
  differs from some other usages; our usage is 
  defined in \SSS\ref{s-formulae-practical}.
The} results are presented in 
\SSS\ref{s-results-fvir}. 
{An {\sc octave} script for making the
calculations and plots is provided in the source package for the 
preprint version of this paper\footnote{\url{http://arXiv.org/abs/1303.4444}}.}
Conclusions are given in 
\SSS\ref{s-conclu}. 

%\section{Method}

\tNbody

\section{\protect The virialisation approximation} \label{s-method-fvir}

\subsection{$N$-body {simulations}} 
\label{s-N-body}

``Dark'' (i.e. baryonic plus non-baryonic) matter halo 
merger history trees have
been calculated since 1992 from $N$-body simulations 
\citep{1993ASPC...51..298R,RPQR97}
and by semi-analytical methods \citep{LaceyColeMilano93}.
Estimates of the virialisation fraction are implied by
these calculations, e.g. Table~6 in \citep{RNinin01}.
Here, the $N$-body simulations used are Virgo
Consortium $256^3$-particle $T^3$ (3-torus) simulations 
\citep{Jenkins98Virgo,Thomas98Virgo}\footnote{\protect\url{http://www.mpa-garching.mpg.de/NumCos}} %% WARNING: Note4 assumes no re-ordering of footnotes!!!
for an EdS-CDM model with
$h=0.5$, 
and normalization in the mean density fluctuation $\sigma_8 = 0.51$
at 8{\hMpc},
{where the Hubble constant, i.e. 
  the zero redshift Hubble parameter,
  is written $H_0 \equiv 100 h$~km/s/Mpc.}
One simulation has comoving box size 239.5~{\hMpc}
and mass per dark matter particle (implicitly, baryonic and non-baryonic together)
of $m = 2.27\times 10^{11} h^{-1} M_\odot$,
the other 84.55~{\hMpc} and 
$m = 1.0 \times 10^{10} h^{-1} M_\odot$.
{The length scale of the latter simulation is small, 
so the simulation is only used for comparison in
Fig.~\ref{f-DEvsvirial}.}
A friends-of-friend group finder {\sc fof-1.1}\footnote{\protect\href{http://www-hpcc.astro.washington.edu/tools/fof.html}{{\tt http://www-hpcc.astro.washington.edu/tools/}}\\ \protect\href{http://www-hpcc.astro.washington.edu/tools/fof.html}{{\tt fof.html}}} at $0.2$ times the mean 
interparticle separation, for a minimum of $8$ particles per group,
i.e. $1.8\times 10^{12} h^{-1} M_\odot$, is used
independently at each output redshift $z$ 
to detect virialised objects at that redshift,
giving the number of
virialised particles $N_{\mathrm{vir}}$ and the complement, i.e. the
number of non-virialised particles $N - N_{\mathrm{vir}}$, where
$N=256^3$ is the total number of particles. 
A different group finder, such as a 
spherical overdensity (SO) group finder, would give somewhat different
results to those found here 
\citep{Tinker08FOFvsSO,MWhite01defnhalo,MWhite02FOFvsSO}, but
since we are interested in the total virialised mass, the
dilemma of whether to define a slightly overlapping pair of haloes as
a single halo (FOF) or a pair of dynamically distinct haloes (SO)
is {insignificant} for the present work.
We define the virialisation fraction 
\begin{equation}
  \fvir(z) := \frac{N_{\mathrm{vir}}(z)}{N}.
  \label{e-defn-fvir}
\end{equation}
{We spline interpolate between \postrefereechangesII{the} simulation
output times 
\postrefereechangesII{shown in Table~\ref{t-N-body}.
  A similar quantity is defined
  in other multi-scale models, e.g. the ``wall fraction'' $f_{\mathrm w}$ in 
  (9) of \cite{DuleyWilt13}, which gives a similar value
  of $\fvir(0)$ (cf Table~\ref{t-N-body} here, Table~1 of 
  \cite{DuleyWilt13}.)}

\subsection{Effective cosmological parameters}

The effective, i.e. volume-averaged, matter density parameter
combines the matter density parameter in the non-virialised region with
that in the (at late times) much tinier virialised region. 
If we think of the particles in their original comoving positions in the background model, then
the volumes occupied by the two regions are 
approximately in the ratio {$(1-\fvir) : \fvir$}.
{However}, taking into account the actual situation of a virialised region, we notice that
the volume occupied
by the virialised matter has shrunk by a factor of about $\deltavir$
(\ref{e-deltavir-EdS}), 
so in the homogeneous model (with no local nor comoving global 
curvature changes), the volume ratio
increases to {$(1 -\fvir/\deltavir) : \fvir/\deltavir$.}
{Following, e.g., {\cite{BuchCarf08,WiegBuch10},}
let us label the non-virialised and virialised regions 
$\cal E$ (for ``empty'') 
and $\cal M$ (for ``massive''), respectively, 
and their disjoint union $\cal D := \cal M \cup \cal E$.
We now establish a general formalism in \SSS\ref{s-formulae-general},
derive formulae for the dark-energy--free, stable clustering case in \SSS\ref{s-formulae-DEfree},
and describe how to use these in \SSS\ref{s-formulae-practical}.}

\subsubsection{General formulae} \label{s-formulae-general}

{Writing $| \CF  |$ for the volume of a 
spatial region $\CF$, let us define}
\begin{equation}
  \postrefereechangesII{\lambda_{\CM}:=
    \frac{\left|\CM\right|}{\left|\CD\right|}.}
\end{equation} 
{This corresponds} to the virialisation volume fraction, {with} 
$\lambda_{\CM} {:=} f_{\rm vir}/\delta_{\rm vir}$.
{As in (25) of \cite{BuchCarf08},
we} then have the linear {combinations}
\begin{eqnarray}
  H_{\CD} &=& \lambda_{\CM}\, H_{\CM} + \left(1-\lambda_{\CM}\right)H_{\CE}
  \label{e-HD}
  \\
  \average{\rho} &=& \lambda_{\CM}\, \averageM{\rho} + \left(1-\lambda_{\CM}\right)\averageE{\rho} 
  \label{e-rhoD}
  \\
  \average{\CR} &=& \lambda_{\CM}\, \averageM{\CR} + \left(1-\lambda_{\CM}\right)\averageE{\CR}\;. 
  \label{e-RD}
\end{eqnarray}
{As shown in \cite{BuchCarf08}, 
\postrefereechangesI{the} volume-averaged Hamiltonian constraint,
\begin{equation}
  3H_{\CD}^{2} =  8\pi G\average{\varrho}-\frac{1}{2}\average{\CR}-\frac{1}{2}\CQ_{\CD}+\Lambda \;,
  \label{eq:Hamilton-Mittel}
\end{equation}
leads to the {kinematical} backreaction on $\CD$,}
\begin{eqnarray}
  \CQ_{\CD} & = & \lambda_{\CM}\CQ_{\CM}+\left(1-\lambda_{\CM}\right)\CQ_{\CE}
  \label{e-QD}\\
  &  & +6\lambda_{\CM}\left(1-\lambda_{\CM}\right)\left(H_{\CM}-H_{\CE}\right)^{2}\;.\nonumber 
\end{eqnarray}
\postrefereechangesII{We also define
  the volume-averaged scale factors, 
  \begin{equation}
    a_{\cal D} := \left(\frac{|\CD|}{|\CD_0|}\right)^{1/3}, \;\;
    a_{\cal M} := \left(\frac{|\CM|}{|\CM_0|}\right)^{1/3}, \;\;
    a_{\cal E} := \left(\frac{|\CE|}{|\CE_0|}\right)^{1/3}, 
  \end{equation}
  as in (3) of \cite{WiegBuch10}, but normalise by the present
  epoch (subscript 0) instead of the initial epoch.}

{As in \cite{WiegBuch10},}
we may now define cosmological parameters on $\cal D$
using (\ref{eq:Hamilton-Mittel}).
{Similarly to the FLRW model}, {these are
derived by dividing (\ref{eq:Hamilton-Mittel}) by} $3H_{\CD}^{2}$.
{For a generic spatial domain $\CF$, which may be any 
of $\CD$, $\CM$, or $\CE$, we {define the cosmological
parameters}
\begin{eqnarray}
  \Omm^{\CF} & := & \frac{8\pi G}{3H_{\CD}^{2}}\langle\varrho\rangle_{\CF}\;\;,\;\;
  \Omega_{\Lambda}^{\CF}:=\frac{\Lambda}{3H_{\CD}^{2}}\;\;,
  \nonumber \\
  \Omega_{\CR}^{\CF} & := & -\frac{\langle\CR\rangle_{\CF}}{6H_{\CD}^{2}}\;\;,\;\;
  \Omega_{\CQ}^{\CF}:=-\frac{\CQ_{\CF}}{6H_{\CD}^{2}}\;.
  \label{e-defn-omegas}
\end{eqnarray}
The choice of dividing by $H_{\CD}^{2}$ 
independently of the choice of $\CF$ is 
motivated  by the stable clustering hypothesis \cite{Peebles1980}, according to which 
the averaged expansion rate in $\cal M$ (virialised regions) is zero, i.e.
$H_{\CM} \approx 0$.}

{With} {these} definitions, the {Hamiltonian} 
constraint (\ref{eq:Hamilton-Mittel}) {on $\CF$ is} 
\begin{equation}
  \Omm^{\CF}+\Omega_{\Lambda}^{\CF}+\Omega_{\CR}^{\CF}+\Omega_{\CQ}^{\CF}=\frac{H_{\CF}^{2}}{H_{\CD}^{2}}\;.
  \label{e-Hamilton-HD}
\end{equation}
{Thus, the dimensionless parameters $\Omega$ defined
this way sum to unity on the combined domain $\CD$, but on $\CM$ 
or $\CE$ are only constrained to sum to a non-negative value, 
which is determined by comparing the region's squared expansion rate to 
that of the combined domain $\CD$.}

\subsubsection{Dark-energy--free, stable clustering case}  \label{s-formulae-DEfree}

{Let us assume no dark energy, i.e. $\Lambda=0$, and the stable clustering
hypothesis for the $\CM$ regions, $H_\CM \approx 0$.  
Thus (\ref{e-HD}) becomes
\begin{eqnarray}
  H_{\CD}= \left(1-\lambda_{\CM}\right)H_{\CE}\;.\nonumber \\
  %\average{\varrho}= \lambda_{\CM}\averageM{\varrho}\;;\nonumber \\
  %\average{\CR}= \lambda_{\CM}\averageM{\CR} + \left(1-\lambda_{\CM}\right)\averageE{\CR}\;;\nonumber \\
  %\CQ_{\CD}  =  \lambda_{\CM}\CQ_{\CM}+\left(1-\lambda_{\CM}\right)\CQ_{\CE}+6\lambda_{\CM}\left(1-\lambda_{\CM}\right)H_{\CE}^{2}\;.\nonumber \\
  \label{e-HD-stableclustering}
\end{eqnarray}
Using (\ref{e-HD-stableclustering}), (\ref{e-rhoD}), (\ref{e-RD}), (\ref{e-QD}), and
(\ref{e-defn-omegas})
we find that the average characteristics are related:
\begin{eqnarray}
  \Omm^{\CD} =  \left(1-\lambda_{\CM}\right) \Omm^{\CE} +   \lambda_{\CM} \Omm^{\CM} \;;\nonumber\\
  \Omega_{\CR}^{\CD} = \lambda_{\CM} \Omega_{\CR}^{\CM}+ \left(1-\lambda_{\CM}\right)\Omega_{\CR}^{\CE}\;;\nonumber\\
  \Omega_{\CQ}^{\CD} =  \lambda_{\CM} \Omega_{\CQ}^{\CM}+ \left(1-\lambda_{\CM}\right)\Omega_{\CQ}^{\CE} -\frac{\lambda_{\CM}}{1 - \lambda_\CM}
  \;.\nonumber\\
  \label{e-omega-relations}
\end{eqnarray}
The Hamiltonian constraint in the form (\ref{e-Hamilton-HD}) gives
three independent equilibria,
\begin{eqnarray}
  \Omm^{\CD} + \Omega_{\CR}^{\CD}+\Omega_{\CQ}^{\CD} &=& 1 \;;\nonumber \\
  \Omm^{\CM} + \Omega_{\CR}^{\CM}+\Omega_{\CQ}^{\CM} &=& 0 \;;\nonumber \\
  \Omm^{\CE} + \Omega_{\CR}^{\CE}+\Omega_{\CQ}^{\CE} &=& \frac{1}{\left(1-\lambda_\CM \right)^2}\;.\nonumber\\
  \label{e-equilibria}
\end{eqnarray}
Together, (\ref{e-omega-relations}) and (\ref{e-equilibria}) consist of,
at a given redshift $z$,}
six algebraic equations {with nine} unknowns, 
{where 
the virialisation volume fraction $\lambda_{\CM}(z)$ is 
estimated by calculating $\fvir(z)$ from a background model
(\ref{e-defn-fvir}) and by using the appropriate 
analytical value of $\deltavir$ for that model.
This set of equations should be used along with the void-dominated
expansion rate (\ref{e-HD-stableclustering}),
since $H_\CD$ is used in defining these $\Omega$ parameters.} 
These {formulae} are valid at each redshift.

{To retain generality while simplifying the algebra,} we
introduce the sums 
\begin{equation}
  \Omega_{\CR}^{\CF}+\Omega_{\CQ}^{\CF} =: \Omega_{X}^{\CF}, 
  \label{e-defn-X-matter}
\end{equation}
so that we {can consider 
$X$~matter to represent the deviation of the averaged parameters from
FLRW behavior.
Thus, six variables are reduced to} three, while 
{equations (\ref{e-RD}) and (\ref{e-QD}) 
  become a single equation.  The resulting set of equations is}
\begin{eqnarray}
  \Omm^{\CD} &=&  
  \left(1- \lambda_{\CM}\right) \Omm^{\CE} + \lambda_{\CM} \Omm^{\CM} \;
  \label{e-eqn-summary-ommD} \\
  \Omega_{X}^{\CD} &=&  
  \lambda_{\CM} \Omega_{X}^{\CM}+ \left(1-\lambda_{\CM}\right)\Omega_{X}^{\CE} 
  -\frac{\lambda_{\CM}}{1 - \lambda_\CM}
  \;   \label{e-eqn-summary-omXD} \\
  \Omm^{\CD}+\Omega_{X}^{\CD}&=& 1 \;
  \label{e-eqn-summary-D-Hamilton} \\
  \Omm^{\CM}+\Omega_{X}^{\CM}&=& 0 \;
  \label{e-eqn-summary-M-Hamilton} \\
  \Omm^{\CE} + \Omega_{X}^{\CE}&=& \frac{1}{\left(1-\lambda_\CM \right)^2}\;, 
  \label{e-eqn-summary-E-Hamilton}
\end{eqnarray}
{where (\ref{e-eqn-summary-E-Hamilton}) is redundant, since it is implied by 
(\ref{e-eqn-summary-ommD})--(\ref{e-eqn-summary-M-Hamilton}). Thus, there
are four independent} algebraic relations 
{with six} unknowns, given the virialisation volume
fraction $\lambda_{\CM}$. 

\subsubsection{Background model} \label{s-formulae-practical}

The equations in \SSS\ref{s-formulae-general} could, in principle, be
solved in a background-free way, \postrefereechangesI{i.e. without
  starting from an FLRW (or other) model.  Here, we use a background
  FLRW model (specifically, the EdS model). That is, we assume that at
  high redshifts the FLRW metric is approximately valid, but at low
  redshifts the virialised and non-virialised subdomains can be
  separately modeled as non-linear deviations from the FLRW model
  ``background''. At high redshifts, the large-scale (domain $\CD$)
  average parameters (\SSS\ref{s-formulae-general},
  \SSS\ref{s-formulae-DEfree}) should be approximately the same as the
  FLRW parameters. At low redshifts, the large-scale average
  parameters are not constrained to match the FLRW parameters, i.e.
  cosmological parameters on $\CD$ at low $z$ are not constrained to
  match the background. Thus, at low redshift, the physical meaning of
  the background model is that it is an extrapolation of the
  FLRW metric from the
  pre-virialisation epoch to the virialisation epoch\footnote{\protect\postrefereechangesI{In analogy,
    one might imagine using $f\supbg(z)=5$ as a high-$z$ background
    model for the function $f(z) = 50 \mathrm{e}^{-z} + 5$. Knowledge
    of $f\supbg$ together with other information (analogous to
    perturbation statistics) might help to calculate $f$, but at low $z$, 
    $f\supbg$ is a poor approximation to $f$.}}. 
  This terminology differs from that used in,
  e.g. \cite[][I~para.~6]{GreenWald11} or LTB void models: at low
  redshift, {\em our background metric is not the large-scale average
    metric}.}

\postrefereechangesI{We first use the EdS background model in order}
  to {approximate} the void region matter density parameter
\begin{eqnarray}
  \Omm^{\CE}  = 
  \frac{8\pi G}{3 H_{\CD}^2} \averageE{\rho} % \nonumber \\
  &  {\approx} &
  \frac{8\pi G}{3 H_{\CD}^2} 
  \left(1-\fvir\right) \left<{\rho}\right>\supbg  \nonumber \\
  &  = &
  (1-\fvir)
  \left(\frac{H\supbg}{H_{\CD}}\right)^2 \Omm\supbg\;, 
  \nonumber \\
  \label{e-ommE-practical}
\end{eqnarray}
where $\langle\rho\rangle\supbg, \Omm\supbg$, and $H\supbg$ 
are the FLRW mean density, density parameter, and 
expansion {rate},
respectively.
{Here, we have assumed that
since $\fvir < 1$  and $\deltavir \approx 178$ 
[EdS case, (\ref{e-deltavir-EdS})],  we have
$\lambda_{\CM} \ll 1$, 
so that the {non-virialised} matter occupies
approximately the full volume. The {virialised} matter density parameter
can also be estimated from the background model, 
\begin{eqnarray}
  \Omm^{\CM}  = 
  \frac{8\pi G}{3 H_{\CD}^2} \averageM{\rho} % \nonumber \\
  &  \approx &
  \frac{8\pi G}{3 H_{\CD}^2} 
  \deltavir \left<{\rho}\right>\supbg  \nonumber \\
  &  = &
  \deltavir
  \left(\frac{H\supbg}{H_{\CD}}\right)^2 \Omm\supbg\;. 
  \nonumber \\
  \label{e-ommM-practical}
\end{eqnarray}
Thus, substituting (\ref{e-ommE-practical}) and (\ref{e-ommM-practical}) 
into (\ref{e-eqn-summary-ommD}) gives
\begin{eqnarray}
  \Ommeff(z) := \Omm^{\CD} &\approx &
  \left[1-\frac{\fvir}{\deltavir}(1-\fvir) \right]
  \left(\frac{H\supbg}{\Heff}\right)^2 \Omm\supbg
  \nonumber \\
  &\approx &   \left(\frac{H\supbg}{\Heff}\right)^2 \Omm\supbg\;, 
  \label{e-ommeff-expanded}
\end{eqnarray}
where the $\CD$-averaged parameter
has been relabeled as an effective parameter. For percent-level
accuracy, the first line in (\ref{e-ommeff-expanded}) is 
necessary, though not sufficient.}

{When assuming a background EdS model,
  the effective expansion rate implied by
  the stable clustering hypothesis, i.e. using
  (\ref{e-HD-stableclustering}), combines the background, time-varying
  expansion rate $H\supbg(z)$ with the peculiar expansion rate
  expressed as a velocity difference normalised by a comoving
  spatial separation, $\Hpeculiarcomov(z)$,
  \begin{eqnarray}
    \Heff(z) &:=&  H_{\CD}  \label{e-defn-Heff} \\
    &=& \left(1-\lambda_{\CM}\right)  
    \left[ H\supbg(z) + \Hpeculiarcomov(z) \; 
      \postrefereechangesII{\aeff}^{-1} \right],
    \nonumber
    \label{e-Heff-general}
  \end{eqnarray}
  where the \postrefereechangesII{$\aeff^{-1}$} factor 
  \postrefereechangesII{(again relabelling $\CD$ as ``eff'')}
  converts the 
  {peculiar expansion rate} from
  comoving to physical length units.
  At redshifts prior to the virialisation epoch, 
  i.e. for $z \agt 3$,
  we must have
  \begin{equation}
    \Heff(z) \approx H\supbg(z).
    \label{e-Heff-hi-z}
  \end{equation}}
This is given by
  the homogeneous Friedmann equation,
\begin{equation}
  H\supbg(z) = \Hbg \sqrt{\Omega_{\Lambda0}\supbg 
    + \Omega_{\mathrm{k}0}\supbg a^{-2} 
    + \Ommzero\supbg a^{-3}},
  \label{e-H-FLRW-general}
\end{equation}
{for} background {model} zero-redshift 
parameters including (in general) the 
Hubble constant {($\Hbg$), and the
dark energy ($\Omega_{\Lambda0}\supbg$),
curvature ($\Omega_{\mathrm{k}0}\supbg$), and 
matter density ($\Ommzero\supbg$)} parameters. In a comoving-rigid
void model (i.e. a standard FLRW model), these background parameters
correspond to zero-redshift parameters. In the virialisation approximation,
these parameters only exist as hypothetical 
projections from high redshift to an idealised
low-redshift state; they are not physically realised.
In the present work, we only consider an EdS
background model, 
\postrefereechangesII{which we extrapolate from high redshift to low,
  i.e. using the effective scale factor $\aeff(t)$ instead of the
  FLRW scale factor $a(t)$,}
so (\ref{e-H-FLRW-general}) simplifies to
\begin{equation}
  H\supbg(z) = \Hbg\,  \postrefereechangesII{\aeff}^{-3/2}.
  \label{e-H-EdS}
\end{equation} 
In order 
to match low-redshift estimates, 
the background model Hubble constant $\Hbg$ should be set 
so that the virialisation approximation zero-redshift 
{value
\begin{eqnarray}
  \Heff(0) &=& 
  { \left[ 1 - \frac{\fvir(0)}{\deltavir} \right] 
    \left[ \Hbg + \Hpeculiarcomov(0) \right]}
  \nonumber \\
  \label{e-Hbg-assumption}
\end{eqnarray}
[cf. (\ref{e-defn-Heff}) and (\ref{e-H-EdS})] matches recent low
redshift estimates of $H_0$ (\SSS\ref{s-obs-inputs}).}

Estimating $\Hpeculiarcomov(z)$ from background
$N$-body simulations would be difficult with standard $T^3$ FLRW
simulations, since the average velocity difference is calculated
along a straight closed curve $\gamma$ across the simulation box,
i.e. it is zero by construction. Thus, here, we estimate
$\Hpeculiarcomov(0)$ from observations (\SSS\ref{s-obs-inputs}).
Determining the redshift dependence 
$\Hpeculiarcomov(z)$ directly from observations would be more
difficult and model-dependent. However, we have two known
constraints: (\ref{e-Heff-hi-z}) and (\ref{e-Hbg-assumption}) must be
satisfied at the high and low redshift limits, respectively.
{At the high redshift limit,} 
a smooth transition at high redshifts seems physically
reasonable. 
{Moreover, 
virialisation cannot occur without reducing the matter density
in the voids. As 
the voids become more and more empty, the imbalance in 
gravitational potential between the centre and edge
of a void becomes stronger and stronger.
Thus, it is physically likely that
$\Hpeculiarcomov(z)$ increases monotonically as $\fvir(z)$ increases
(i.e. as $z$ decreases).
Here,} we propose a functional form for $\Hpeculiarcomov(z)$ proportional
to the virialisation fraction, i.e.
\begin{equation}
  \Hpeculiarcomov(z) = \Hpeculiarcomov(0) \, \frac{\fvir(z)}{\fvir(0)}.
  \label{e-Hpec-proposal}
\end{equation}
By definition, this formula satisfies the high and low redshift limits,
{has a smooth high-redshift transition, 
and has $\Hpeculiarcomov(z)$ increasing monotonically as $\fvir(z)$
increases. Checking the detailed shape of this function should be considered
as both an observational and a theoretical test of the hypothesis that
the EdS model is the correct background model. Neither task is trivial,
however, since both need to be done in the framework of the scalar
averaging approach
\prerefereechanges{(e.g., using the relativistic Zel'dovich approximation \citep{BuchRZA1,BuchRZA2})}, 
rather than in the perturbed FLRW framework.}

{Since we assume zero dark energy in our
background model and we approximate $\OmQD(z)$ as small, 
(\ref{e-eqn-summary-D-Hamilton}),
(\ref{e-hamiltonconstraint}), and
(\ref{e-defn-X-matter}) give} 
the effective sign-reversed curvature parameter
\begin{equation}
  \OmReff(z) = 1- \Ommeff(z),
  \label{e-defn-omkeff}
\end{equation}
{where 
$\CD$-averaged parameters are again relabeled as effective parameters.}
The
effective comoving 
curvature radius at a given 
epoch can now be written
\begin{equation}
  \RCeff(z) = \frac{c}{\postrefereechangesII{\aeff} \; \Heff(z)} \frac{1}{\sqrt{\OmReff(z)}},
  \label{e-defn-RCeff}
\end{equation}
where for simplicity, we assume that 
{$\OmReff > 0 \; \forall z$}.

{Using
(\ref{e-H-EdS}), 
we can rewrite 
{the first line of
(\ref{e-ommeff-expanded}) for the EdS case 
as
\begin{eqnarray}
  \Ommeff(z) 
  &\approx& 
  \left[1-\frac{\fvir}{\deltavir}(1-\fvir) \right]
  \left(\frac{\Hbg}{\Heff}\right)^2 \; 
  \postrefereechangesII{\aeff}^{-3} , \nonumber \\
  \label{e-ommeff-approx}
\end{eqnarray}
i.e.,} an effective matter density parameter that is
clearly less than unity during the virialisation epoch.}

\subsection{Observational assumptions} \label{s-obs-inputs}

Two low-redshift limit observational estimates are needed
in order to evaluate (\ref{e-defn-Heff}). We adopt 
\begin{eqnarray}
  \Heff(0) &=& 74.0\pm 1.6 \,\mathrm{km/s/Mpc}  
  \label{e-Hbg-estimate}
\end{eqnarray}
using the recent low-redshift Hubble constant estimates
of $H_0 = 73.8\pm 2.4$~km/s/Mpc \citep{Riess11H74} 
and $H_0 = 74.3\pm 2.1$~km/s/Mpc 
\citep{Freedman12H74}.

{To estimate the 
present value of the {peculiar expansion rate} 
$\Hpeculiarcomov(0)$, we divide the typical infall velocity 
$\vinfall$
around rich clusters of galaxies by the typical void radius
$\Dvoid/2$, where both are for low-redshift data
\postrefereechangesII{for roughly comparable redshift limits
  and numbers of objects.}
Out to a distance of 130{\hMpc} 
and at 10 or more degrees above the Galactic Plane,
the median size of eight voids found in 
the IRAS Point Source Catalog redshift survey (IRAS/PSCz) 
is $\Dvoid/2 = 28.3$~{\hMpc} \citep{PlionisIRASPSCz02}.
In Data Release 5 of the Sloan Digital Sky Survey (SDSS), 
a much larger sample of 222 voids found in the redshift range
$0.04 < z < 0.16$ in a solid angle of about 2300 deg$^2$ 
has a mean comoving void radius of 
\begin{equation}
  \Dvoid/2 \approx 25 \pm 2~h^{-1}~\mathrm{Mpc}
  \label{e-obs-dvoid}
\end{equation}
(\citep{FosterNelsonSDSSvoids09}; standard error
by inspection of Fig.~4).}

\postrefereechangesII{A lower redshift analysis, to $z \alt 0.025$ and 
$z \alt 0.028$,
finds 19 and 35 voids 
in the updated Zwicky Catalog and the IRAS Point Source Catalog 
redshift survey, respectively \cite{HoyleVog02voids}, 
with similar 
average effective estimates of $\Dvoid/2 \approx 15 \pm 1.8${\hMpc}.
An analysis of Data Release 7 of the SDSS volume-limited to $z=0.107$ 
finds 1054 voids, with a median effective void radius of
$17${\hMpc} \cite{Pan2012voids}. In addition to the differences in 
redshift and apparent magnitude limits and numbers of voids found, 
observational void analyses vary depending on 
algorithmic details and definitions of void overlap and alignment.}

{Infall 
velocities are not normally derived from observations with the
aim of estimating $\Hpeculiarcomov$. 
The Cluster and Infall Region Nearby Survey (CAIRNS)
study of nine low-redshift rich clusters gives a caustic
outline for infall velocities in front and behind the
clusters of about
\begin{equation}
  \vinfall \approx 2 \sigma_v,
  \label{e-vinfall-vs-sigmav}
\end{equation}
where $\sigma_v$ is the line-of-sight velocity dispersion of cluster
galaxies (Fig.~4, \citep{CAIRNS03I}, within one sky-plane virial
radius of the cluster centres).  {See \citep{Kaiser87} for a
  discussion of redshift space effects interpreted using a homogeneous
  model, especially Fig.~5 illustrating caustics related to the
  turnaround radius.}  Velocity dispersions of 91 rich clusters of the
ESO Nearby Abell Cluster Survey (ENACS) and earlier surveys selected
out to $z=0.1$ from a solid angle of $\approx 8400$~deg$^2$ have a
mean of $\sigma_v = 642 \pm 24$~km/s (Table 1a,
\citep{Mazure96sigmav}; standard error in the mean; cf Fig. 5a of
\citep{Mazure96sigmav}).  An SDSS Data Release 4 analysis of 109
clusters out to $z < 0.1$ over $6700$~deg$^2$ yields $\sigma_v = 585
\pm 15$~km/s (Table 1,
\citep{Aguerri07sigmav}\footnote{\protect\postrefereechangesII{Table 1
    is absent from arXiv:0704.1579v1 (which incorrectly enumerates
    Table~2), but available
    at\\ %% EDITOR: feel free to remove this line break if convenient
    \protect\url{http://vizier.u-strasbg.fr/viz-bin/VizieR?-source=J/A+A/471/17}.}};
standard error in the mean).  The two catalogues have very few members
in common, so an inverse-variance weighted mean can be calculated:
\begin{equation}
  \sigma_v = 601\pm 13 \mathrm{km/s}.
  \label{e-sigmav}
\end{equation}

\postrefereechangesII{The estimates in (\ref{e-obs-dvoid}) and 
(\ref{e-sigmav}) approximately correspond to matching void and cluster 
observational analyses, for $z \alt 0.1$ and about 100--200 
voids or clusters.} 
Thus, with $h=0.74$ from (\ref{e-Hbg-estimate}) and using 
(\ref{e-obs-dvoid}), 
(\ref{e-vinfall-vs-sigmav}), and
(\ref{e-sigmav})
we set
\begin{eqnarray}
  \Hpeculiarcomov(0) := \frac{2\vinfall}{\Dvoid} 
  &\approx & \frac{4\sigma_v}{\Dvoid}
  \label{e-obs-Hpeccomov}
  \\
  &= & 36 \pm 3 \;\mathrm{km/s/Mpc}.
  \nonumber 
\end{eqnarray}}
\postrefereechangesII{This error estimate only includes the 
  random error associated with the observational analyses
  indicated above. Systematic error due to differing catalogue
  limits and void definitions would lead to a larger total error.}

\subsection{Effective metric} \label{s-eff-metric}

\postrefereechangesII{Given the above expressions for $\Ommeff(z)$
  (\ref{e-ommeff-approx}), $\OmReff(z)$ (\ref{e-defn-omkeff}), and
  $\Heff(z)$ (\ref{e-Heff-general}), (\ref{e-H-EdS}) and
  (\ref{e-Hpec-proposal}), the early spatial sections corresponding to
  an EdS model must evolve to spatial sections that are negatively
  curved. To preserve large-scale, statistical, spatial homogeneity
  during the virialisation epoch, we adopt an effective metric that is
  homogeneous and hyperbolic on each spatial section, using the
  effective curvature of that epoch. That is, 
  following \cite{Paranjape08,Larena09template},  we extend one of
  the common expressions of the FLRW metric to an effective,
  spherically symmetric, comoving, observer-centred expression of an
  effective metric with an effective scale factor $\aeff(t)$:}
\begin{eqnarray}
  \diffd s^2 &=& -c^2 \diffd t^2 + %\nonumber \\  &&
  \postrefereechangesII{\aeff}^2(t) 
  \left[
  \diffd {\chieff}^2 +
  {\RCeff}^2 \sinh^2 \frac{\chieff}{\RCeff}  
  \left(\diffd \theta^2 + \cos^2 \theta \, \diffd \phi^2 \right) \right],
  \nonumber \\
  \label{e-metric-eff}
\end{eqnarray}
where the differential radial comoving distance is
\begin{equation}
  \diffd \chieff(z) = 
  \postrefereechangesII{
  \frac{c\, \diffd t}{\aeff} = 
  \frac{c\, \diffd t}{\aeff\, \diffd \aeff} \, \diffd \aeff = 
  \frac{c}{\aeff^2 \; \Heff(z)} \; \diffd \aeff,}
  \label{e-dchieff}
\end{equation}
and $c$ is the conversion constant between space and time units.
Equation (\ref{e-dchieff}) can be integrated numerically to obtain
$\chieff$ and the luminosity distance 
\begin{equation}
    \dLeff(z) = 
    (1+z) \RCeff \sinh \frac{\chieff}{\RCeff} .
    \label{e-dL-eff}
\end{equation}
The effective differential volume element per square degree, necessary
for faint galaxy number count analyses, is
\begin{equation}
  \dVdzeffmath(z) = \left(\frac{\pi}{180}\right)^2
  {\RCeff}^2 \sinh^2 \frac{\chieff}{\RCeff} \frac{\diffd \chieff}{\diffd z}. 
  \label{e-dVdz-eff}
\end{equation}

%\fvdiffrms

The luminosity distance for the virialisation approximation and for the 
homogeneous EdS and $\Lambda$CDM models can be normalised to
the Milne model for convenience, yielding distance moduli $m-M$.
The fraction of the distance modulus that would be needed to correct the
homogeneous EdS model to match the $\Lambda$CDM model can be
written as follows:
\begin{equation}
  \fracdistmod :=  
  \frac{\log_{10} \dLeff - \log_{10} d_{L}^{\mathrm{EdS}}}
       {\log_{10} d_{L}^{\Lambda\mathrm{CDM}} 
         - \log_{10} d_{L}^{\mathrm{EdS}}}.
       %%  (log10(dLeff_iter(2:nz)).-log10(dLEdS(2:nz))) ./ \
       %%        (log10(dLLCDM(2:nz)).-log10(dLEdS(2:nz)));
       \label{e-defn-fracdistmod}
\end{equation}

\fommeff

\fHeff

\fRCeff

\fmetricVF

\fmetricVFfraction

\fmetricVFdVdz

\section{Results}
\label{s-results-fvir}

The virialisation fraction $\fvir$ from the two $N$-body simulations
has already been shown in Fig.~\ref{f-DEvsvirial}.  
{Figure~\ref{f-ommeff}} shows that, as expected,
the effective matter density parameter contributing to curvature drops
as the redshift decreases to zero. The low-redshift value, 
{$\Ommeff(0) = 0.26 \pm 0.05$,} \postrefereechangesII{is} close to observational low-redshift estimates
of $\Omm \sim 0.3$ 
\postrefereechangesII{\cite[e.g.][]{FeldmanJuszk03}}. It is {\em not} a result of fitting the
virialisation approximation to observational data. 
Apart from assuming
an EdS background model, the only observational 
{values used in the
formulae above are those presented in \SSS\ref{s-obs-inputs}, 
i.e. $\Heff(0)$ (\ref{e-Hbg-assumption}) and $\Hpeculiarcomov(0)$ (\ref{e-obs-Hpeccomov}).}
\postrefereechangesII{Systematic error can be roughly estimated as follows.
  If the $\Dvoid/2 \approx 17${\hMpc} 
  estimate of \cite{Pan2012voids} for 1054 voids were used 
  together with the $\sim$100-cluster estimate of
  $\sigma_v$ given in (\ref{e-sigmav}), 
  i.e. without correspondingly
  using a velocity dispersion typical of more common, 
  less massive clusters, then (\ref{e-obs-Hpeccomov}) would
  give $\Hpeculiarcomov(0) \approx 52 \pm 6$~km/s/Mpc
  and $\Ommeff(0) = 0.09 \pm 0.05$, i.e., 
  recent growth in negative curvature would
  be stronger than for our best estimate. 
  If, instead, the low-redshift estimate $H_0 = 63.7 \pm 2.3$~km/s/Mpc \cite{Tammann13}
  %%   0.23290    0.18960   0.15076
  were used or $\sigma_v$ were arbitrarily increased to $\sigma_v = 1000$~km/s,
  %%(sometimes cited informally as a ``typical'' velocity dispersion of a massive cluster),
  %% 0.066739 0.036402 0.015190 
  then $\Ommeff(0) = 0.19 \pm 0.04$ or
  $\Ommeff(0) = 0.036^{+0.03}_{-0.02}$ would be inferred,
  respectively, again giving stronger growth in negative curvature.
  (The $H_0 = 75.4 \pm 2.9$ estimate of \cite[][final version]{RiessFVG12} would
  give $\Ommeff(0) = 0.27 \pm 0.04$.)
  To obtain a {\em weaker} growth in negative curvature for
  increasing cosmological time $t$, either a higher typical void size,
  a lower cluster velocity dispersion, or a higher low-redshift $H_0$
  estimate than our literature values [(\ref{e-obs-dvoid}),
    (\ref{e-sigmav}), (\ref{e-Hbg-estimate}), respectively] 
  would be needed.}

Figure~\ref{f-Heff} shows the effective expansion rate evolution,
which, along with $\fvir$, explains the lower effective matter density
parameter at low redshifts. Through the comoving curvature radius 
(\ref{e-defn-RCeff}), the void-dominated expansion rate at low
redshifts also leads to stronger negative average 
curvature, by decreasing $\RCeff$.
Figure~\ref{f-RCeff} shows the effective curvature radius
evolution $\RCeff(z)$.

Figure~\ref{f-metricVF} 
shows that applying the virialisation 
approximation to the EdS model brings the EdS distance modulus 
{close} to the (homogeneous) $\Lambda$CDM
distance modulus. 
The fraction $\fracdistmod$ (\ref{e-defn-fracdistmod})
provided by the virialisation approximation is shown
in Fig.~\ref{f-metricVFfraction}, 
{showing that a discrepancy remains for $z < 1$.}
{Figure~\ref{f-metricVFdVdz} shows that 
the volume element needed to match faint number counts is} 
provided by this approximation.

\section{Discussion} \label{s-disc}

The dominance of negative curvature over positive curvature on large
scales at recent
epochs has already been postulated based on detailed kinematical and
curvature backreaction estimates \cite{BuchCarf08} and models 
\citep{rasanenFOCUS,Buchert11Towards,buchertrasanen,wiltshireFOCUS,BuchRZA2,Larena09template,RoyBuch10chaplygin,Clarkson11backreaction,WiegBuch10}.  
By definition, over-and under-densities of comoving perturbations in an FLRW background model average
to zero. In general, this is also true on mass-preserving Lagrangian domains $\cal D$ in the nonlinear regime. 
The restmass conservation law assures compensation to hold 
\prerefereechanges{throughout} the evolution of structure.
However, the intrinsic curvature does not obey a conservation law: 
{only a dynamical relation between 
the kinematical backreaction, curvature, and volume evolution}
is conserved \cite{BuchCarf08},
as
{is used}
in this paper through {(\ref{e-hamiltonconstraint}) with
$\Omega_\Lambda=0$;} see the discussion in \cite{buchertrasanen}.

Virialised regions are, by definition, highly
concentrated, occupying very little volume. 
{Thus,} 
while the average {{of} the density deviations from the background} tends to zero, 
the average peculiar curvature (average curvature with 
respect to the background FLRW model curvature)
tends to a negative value when a high fraction of matter has
virialised.
The key formula in \cite{BuchCarf08} is (104), with
a general discussion in Section~5 of strategies for calculating
a more detailed approximation than the one presented here.

At small enough scales, 
{the backreaction model based on a volume average of} 
the relativistic Zel'dovich approximation
mentioned above, representing the effects of the
statistical formation of structure in detail
\citep{BuchRZA1,BuchRZA2}, shows that positive curvature dominates in collapsing regions.
It is speculated that, on these smaller scales, 
positive curvature
plays the role that is usually attributed to ``dark matter'', i.e.
that a substantial portion of dark matter might also be an artefact
of using a Newtonian approximation to structure formation.
{Although 
{the estimates in this paper} deal with void scales and larger,
detailed relativistic calculations would have to take this into account. We do not
attempt to model this here, since {we consider our results 
to be an initial quantitative estimate
of an important physical effect, suggesting that it is worthwhile to go beyond the model presented here.}}

{For over a decade, radially
  inhomogeneous models, both Stephani \citep{DabHend98}
  and 
  \postrefereechangesI{LTB \citep{MustaphaHE97,Celer99,Biswas10}} 
  models, have been
  proposed \postrefereechangesI{(and rejected \citep[e.g.][]{MossZS11defeat})} 
  as dark-energy--free, relativistic fits to
  the type Ia supernovae 
  {luminosity-distance--redshift} relation,
  and have become best known as ``void models''
  \citep{Tomita01void,BCK11review} on a scale of up to $\sim
  2.5$~Gpc (e.g., \citep{GBH08Gpc2p5}), although ``hump
  models'' have also been inferred from the data
  \citep{CBK10,KolbLamb09}.  There has been much debate
  over whether or not the fits should be excluded on
  either observational {grounds
    or as being in
    conflict with the Cosmological Principle.}}
{(See, e.g., \cite{KB12lightdriftanddebunking}.)}

{The virialisation approximation
  (which is neither a Stephani nor an LTB model)
  suggests another interpretation that may resolve some issues 
  in these debates.  
  It is true that in
  a comoving, constant time slice at the present epoch
  $t_0$, a reasonable projection of our past {light} cone
  observations forward to $t_0$ is hard to
  reconcile---at least for large-scale homogeneous
  models---with our location near the centre of a void
  as large as a gigaparsec.  
  {However, cosmological observations are not 
    made on the $t_0$ time slice.}
  {\em On the past
    {light} cone}, the virialisation epoch approximately
  corresponds to an observer-centred ``void'' of a
  gigaparsec or so in size, where ``void'' here means,
  e.g., the spherical region around the observer
  extending over the redshift range where $\fvir(z) \agt
  0.1$. The ``void'' that is relevant for an effective
  metric is that defined by the volume-averaged density
  parameter on our past {light} cone, and this ``void'' is
  mirrored by virialisation. We are located at a highly
  privileged, highly non-random, centralised position on
  the past {light} cone: at the apex of the cone, which is
  the epoch of highest virialisation.}
{Thinking spatially at constant $t$, 
  locating the observer 
  at the centre of a (non-averaged) void 
  is a hypothesis,
  while on the past {light} cone, 
  our privileged position is a geometrical fact.}

{With hindsight, 
\postrefereechangesI{some of}
the {\em numerical} results
of LTB void models by, e.g., \citep{Alnes05voidmodel} and 
the LTB void model $H_0(r)$ estimates of \citep{Enqvist08LTBHpec},
for a radial coordinate $r$ at constant cosmological time,
roughly correspond to the results found here, but with fundamental
differences in their derivation and interpretation
\postrefereechangesI{(e.g. the notion of a background FLRW model).}
 Here, we start
with a large-scale homogeneous (EdS) model and use standard $N$-body
simulations to 
{quantify an approximate, effective} metric 
{at low redshifts, providing a general-relativistic
correction to the FLRW metric.}
{Our} resulting ``void'' is a {past-light-cone}, 
{virialisation-epoch,} volume-averaged pseudo-void only.
In the LTB models, $H_0$ and $\Ommzero$ vary with the radial coordinate
$r$ at a constant cosmological time $t$, 
{at which there is} a true void rather
than a virialisation-epoch ``void''. However, the LTB model parameters are motivated by and interpreted
in relation to observations---on the past {light} cone. Thus, it is
unsurprising that 
\postrefereechangesI{some of} 
the numerical results are \postrefereechangesI{roughly
similar to ours. Nevertheless, constant time-slice voids
appear to require small sizes
(200--250~{\hMpc}; \cite{AlexBiswas09void250Mpc}),
since there are observational difficulties for 
1~{\hGpc}-scale constant time-slice voids 
(e.g. the kinetic Sunyaev-Zel'dovich effect 
\citep{ZhStebbins11void1Gpc,ZibinMoss11voidkSZ}).}

\postrefereechangesII{A much more closely related model is the
  timescape model
  \citep{Wiltshire07clocks,Wiltshire07exact,Wiltshire09timescape,
    SmaleWilt11SNe,Wiltshire12Hflow,DuleyWilt13}.  This model also
  uses virialised and non-virialised domains and scalar averaging, but
  distinguishes between the parameters estimated by an observer in a
  randomly chosen position from those made by an observer located in a
  virialised object, using a {\em phenomenological lapse
    function}. The estimates in the present paper should most closely
  correspond to, e.g., the ``bare'' (volume-averaged for a volume-random
  observer; 
  see \cite{BuchCarf03nudity} for the original usage of this term)
  parameters of Table~1 in the most recent timescape
  calculations \cite{DuleyWilt13}.
  In our case, we have matched $\Heff(0)$ to the usual 
  low-redshift observational estimates. This is about 20~km/s/Mpc
  greater than $\bar{H}_0$ of  Table~1 of \cite{DuleyWilt13},
  and our age of the Universe estimate is, unsurprisingly, about 5~Gyr 
  less than that for the volume-average observer, $t_0$, of Table~1 of \cite{DuleyWilt13}.
  Our zero-redshift matter density is about 50\%
  higher than the timescape value.
  Since we are located in a galaxy, it is clear that 
  a correction to take into account our non-random location
  will be needed in future work on the present approach. 
  Nevertheless, the recent timescape calculation shares many 
  similarities with ours, in particular agreeing
  with ours in that the volume-averaged curvature is
  strongly negative at the present epoch.}
\postrefereechangesII{The recent Swiss cheese/LTB approach of
  \cite{Lavinto13} (see also references therein) also agrees with
  ours in the sense that it finds that inhomogeneity can lead
  to strongly negative curvature and a significantly increased 
  expansion rate.}

\postrefereechangesI{A more distantly related family of inhomogeneous,
  relativistic models is that of expanding vacuum solutions containing
  regularly spaced black holes
  \citep{Clifton12BH,Yoo12BH,Clifton13BH,Bentivegna13lattice}. These
  are potentially interesting for studying topological acceleration
  \citep{RBBSJ06,RR09,ORB11}, but do not represent the transition
  epoch from pre-virialisation to virialisation.}

\fmetricVFqdecel

Is the scale factor accelerating according to 
{the virialisation} approximation?
{The study of LTB models shows that for
  inhomogeneous universe models in general, at least two different
  definitions of the deceleration parameter can be usefully made, and
  luminosity-distance--redshift observations do not imply model-free
  estimates of either of these \citep{Bolejko08accel}.
Here, using the path of a radial photon 
in (\ref{e-metric-eff}),
$  \diffd \chieff/ \diffd a$ from (\ref{e-dchieff}),
and appropriately converting between space
and time units, an effective deceleration 
{parameter,}
\begin{equation}
  \qdeceleff(z) := 
  \postrefereechangesII{-\frac{\ddotaeff \aeff}{\dotaeff^2},}
  \label{e-defn-q-decel}
\end{equation}
can be defined.
\postrefereechangesII{As can be expected from Fig.~\ref{f-Heff},
  $\qdeceleff$ decreases as $z$ increases.
  Figure~\ref{f-metricVFqdecel} shows that for 
  the definition in (\ref{e-defn-q-decel}),
  moderate acceleration occurs for $z\alt 1.5$. 
  The change from deceleration, during the early (EdS) phase, to
  acceleration at a late epoch occurs when $\ddotaeff = 0$, i.e., when
\begin{equation}
  \frac{\dotaeff}{\aeff} \frac{\diffd \aeff}{\diffd \chieff} 
  = \frac{\diffd^2 \aeff}{\diffd t \,\diffd \chieff}.
  \label{e-accel-switch}
\end{equation}}

{Given that only two observational values are
  used in the above approximation, 
  i.e. one that is commonly derived from observations, 
  $\Heff(0)$ (\ref{e-Hbg-assumption}), and 
  one that we derive here from observations, 
  $\Hpeculiarcomov(0)$ (\ref{e-obs-Hpeccomov}),
  it would be \postrefereechangesII{good} to see if the approximation agreed with other
  further observational constraints beyond
  $\Ommeff(0) = 0.26 \pm 0.05$
  and the $\dLeff(z)$ relation.
  One important constraint is the age of the Universe
  at the (approximately comoving) observer's spacetime location.
  Using the EdS age at $z=10$,
  the virialisation approximation gives the present-day age of the Universe 
  \postrefereechangesII{$t_0 = 12.7 \pm 0.6$~Gyr.}
  This is a little low, but consistent with 
  \prerefereechanges{$t_0 = 13^{+4}_{-2}$~Gyr} from the
  stellar population dating of moderate
  redshift elliptical galaxies over $0.3 < z < 0.9$
  \citep{Ferr01ellipticalage},
  or \prerefereechanges{$t_0 \agt 12.8 \pm 1$~Gyr} from globular cluster ages
  calibrated using Hipparcos parallaxes
  \citep{KraussGC00}.
\postrefereechangesII{Ideally, it would best to estimate $\Hpeculiarcomov(0)$ 
  using both cluster velocity dispersions and void sizes from a single
  observational volume.}

\section{Conclusion} \label{s-conclu}

The approximation presented above assumes ``old physics'' (general
relativity), not ``new physics''. 
It is a rough approximation of old physics, general relativity,
applied to the old observations of a high virialisation fraction 
(galaxies, clusters of galaxies)
at recent epochs and at a small distance scale in the past {light} cone. 
Numerical values of $\fvir(z)$ {are esimated}
from a standard EdS $N$-body simulation 
\citep{Jenkins98Virgo,Thomas98Virgo}.
{Only two observational values are assumed: the
zero-redshift Hubble parameter (\ref{e-Hbg-assumption})
and the zero-redshift 
{peculiar expansion rate} 
across non-virialised regions, $\Hpeculiarcomov(0)$ (\ref{e-obs-Hpeccomov}).
The inferred effective low-redshift matter density parameter
is realistic, $\Ommeff = 0.26 \pm 0.05$ 
\postrefereechangesII{(random error only)}, and the virialisation-corrected EdS distance modulus
is close to the $\Lambda$CDM distance modulus
(Figs~\ref{f-metricVF}, \ref{f-metricVFfraction}).}

Of course, 
virialised objects cannot be completely ignored, the
{virialisation fraction is derived from an}
$N$-body simulations with a fixed
(zero) curvature EdS {model},
and different choices of parameters in the $N$-body analysis 
(group finder algorithm, group finder minimum number of particles,
definition of the void size)
and the $N$-body simulation itself
(calculation algorithm, volume of the simulation, particle mass 
resolution) would be likely to modify the above calculations.
Our proposed interpolation 
(\ref{e-Hpec-proposal}) and normalisation
(\ref{e-obs-Hpeccomov}) 
of $\Hpeculiarcomov(z)$ could also be improved in many ways.}
It is, however, unlikely that these refinements would substantially reduce
the amplitude of the corrections
estimated here, since the observational evidence in
favour of a high virialisation fraction at the present epoch 
$z \ll 1$ and the observational and theoretical evidence for
{$H(z) + \Hpeculiarcomov(z) \; a^{-1} $}
(\ref{e-defn-Heff}) to roughly follow the
values shown in Fig.~\ref{f-Heff} is strong. 
\postrefereechangesII{For example, arbitrarily modifying 
  the ${\fvir(z)}/{\fvir(0)}$ interpolation of (\ref{e-Hpec-proposal}) 
  to either $({\fvir(z)}/{\fvir(0)})^{1/2}$ or
  $({\fvir(z)}/{\fvir(0)})^{1/2}$ would only change our $t_0$ estimate
  to $t_0 = 13.6 \pm 0.8$~Gyr or
  $t_0 = 11.9 \pm 0.4$~Gyr, respectively.}
A more detailed
approximation than that presented here might show that a hyperbolic
or spherical, 
{zero-dark-energy} background FLRW metric,
corrected for virialisation, would better match
{the full range of observational constraints.}
{It may also be important
to consider inhomogeneous light-path effects on standard candles,
such as type Ia supernovae, as a bias when assuming the FLRW
models as a family of background
models}
\citep{KainMarra10lensingbias,Clarkson12lensingbias,FleuryDU13}.
High redshift observations
should be modelled and interpreted in a way that avoids dependence on the low
redshift metric. 
{Low-redshift peculiar-velocity observations
(e.g. \citep{Tully08preHpec}) would be useful if designed to optimally
estimate $\Hpeculiarcomov(0)$.}

Pending more accurate, relativistic
calculations, it would seem prudent to consider ``dark energy'' as an
artefact of virialisation-induced negative spatial curvature 
{and void-dominated expansion rates, both 
of these being physical properties that are}
neglected in the standard cosmological framework.

\bigskip %% shouldn't the acknowledgments environment do this automatically?

\begin{acknowledgments}
Thank you to Martin Kerscher, Tomasz Kazimierczak,
{Alexander Wiegand}, 
Bruce A. Peterson, Bartosz Lew, 
\postrefereechangesI{Hirokazu Fujii,}
\postrefereechangesII{and an anonymous referee}
for useful comments.
Some of this work was carried out within the framework of the European
Associated Laboratory ``Astrophysics Poland-France''. 
\postrefereechangesII{Part of this work consists of research conducted in the scope of the HECOLS International Associated Laboratory.}
BFR thanks
the Centre de Recherche Astrophysique de Lyon for a warm welcome
and scientifically productive hospitality.
A part of this work was conducted within the ``Lyon Institute of
Origins'' under grant ANR-10-LABX-66.
\prerefereechanges{Some of JJO's contributions to this work
were supported by the Polish Ministry of Science and Higher Education
under ``Mobilno\'s\'c Plus II edycja''.}
A part of this project has made use of 
Program Oblicze\'n Wielkich Wyzwa\'n nauki i techniki (POWIEW)
computational resources (grant 87) at the Pozna\'n 
Supercomputing and Networking Center (PCSS).
The $N$-body simulations analyzed in this paper were carried out by
the Virgo Supercomputing Consortium using computers based at the Computing
Centre of the Max-Planck Society in Garching and at the Edinburgh
Parallel Computing 
Centre\footnote{\protect\url{http://www.mpa-garching.mpg.de/NumCos}}.
Use was made 
of 
%the computer algebra program {\sc maxima},
the GNU {\sc Octave} command-line, high-level numerical computation software 
(\url{http://www.gnu.org/software/octave}),
and
%of the WMAP data
%(\url{http://lambda.gsfc.nasa.gov/product/}), 
the
Centre de Donn\'ees astronomiques de Strasbourg 
(\url{http://cdsads.u-strasbg.fr}).
%and the GNU {\sc plotutils} plotting package.
\end{acknowledgments}

%%submit:figures-at-end  %% this is for the APS submission style

%\subm{\clearpage}

%\section*{References}

\providecommand{\href}[2]{#2}\begingroup\raggedright\endgroup
 
%%% CUT_HERE ... BUT LINES BELOW WILL BE INCLUDED - only \end{document} is added afterwards

\end{document}